\documentclass[fleqn,usenatbib]{mnras}

\usepackage{newtxtext,newtxmath}

\usepackage[T1]{fontenc}

\DeclareRobustCommand{\VAN}[3]{#2}
\let\VANthebibliography\thebibliography
\def\thebibliography{\DeclareRobustCommand{\VAN}[3]{##3}\VANthebibliography}

\usepackage{graphicx}	
\usepackage{amsmath}	

\title[K2-18b Moons]{Can Moons Exist around the Habitable-zone Planet K2-18b?}

\author[S. D. Patel et al.]{
Shaan D. Patel,$^{1}$\thanks{E-mail: shaan.patel@uta.edu}
Billy Quarles,$^{2}$\thanks{E-mail: billylquarles@gmail.com}
Manfred Cuntz$^{1}$\thanks{E-mail: cuntz@uta.edu},
and Nevin N. Weinberg$^{1}$\thanks{E-mail: nevin@uta.edu}
\\
$^{1}$Department of Physics, University of Texas at Arlington, Arlington, TX 76019, USA\\
$^{2}$Department of Physics and Astronomy, East Texas A\&M University, Commerce, TX 75428, USA
}

\date{Accepted 2025 July 15. Received 2025 July 11; in original form 2025 June 29}

\pubyear{\the\year{}}

\begin{document}
\label{firstpage}
\pagerange{\pageref{firstpage}--\pageref{lastpage}}
\maketitle

\begin{abstract}
K2-18b closely orbits a nearby M3 dwarf within its habitable zone, where this planet could be either a super-Earth or a mini-Neptune.  Recent studies using transit spectroscopy suggest that it is Hycean in nature, but this classification is currently controversial.  We use the N-body integrator \texttt{rebound} and its extension library \texttt{reboundx} to investigate the possibility of exomoons around K2-18b.  Due to tidal interactions that induce outward migration, we find that any moons would be extremely unlikely.  If formed, their lifetimes would be relatively short, not exceeding 10 Myr assuming Earth-like or Neptune-like tidal parameters for K2-18b.  Recent studies estimate the stellar (and system) lifetime as 3 Gyr, which is significantly longer than the tidal migration timescale. We show that exomoons are unlikely to survive around K2-18b due to rapid tidal-driven migration, casting doubt on moon-based habitability scenarios for short-period M-dwarf planets in general.
\end{abstract}




\section{Introduction}

The star--planet system K2-18 consists of a red dwarf and (at least) two planetary companions in relatively close proximity to Earth; see {\it Gaia} Data Release 3 \citep{gaia23}.  One of those planets is K2-18b, identified as a super-Earth or a mini-Neptune and discovered in 2015 by the {\it Kepler} space telescope in its K2 mission \citep{montet15}.  The host star is of spectral type M3V with a mass of $0.495 \pm 0.004~M_\odot$ and an age of $3.0 \pm 0.1$~Gyr; see \citet{cloutier19} and \citet{sairam25} for details.

K2-18b recently received increased attention as a possible Hycean world in conjunction with the search for habitable environments and biomarkers in exoplanetary atmospheres, revealing the presence of methane, carbon dioxide, and possible hints of dimethyl sulfide (DMS) and dimethyl disulfide (DMDS) \citep{madhus23, madhus25}\footnote{Notably, these findings have led to some controversy in recent works, including a non-biological interpretation of the data; see \citet{luque25} and \citet{welbanks25} for details.}. These molecules are indications of carbon-based chemistry, which could be due to the presence of life, if K2-18b is identified as a super-Earth; however, to a much lower extent if identified as a mini-Neptune. In this work, both planet types will be considered.

Noting that K2-18b is located in the stellar habitable zone (HZ), a potentially important alternate mode for facilitating habitability is the presence of one or more large exomoons hosted by the planet.  However, any moon (if formed) must be orbitally stable for a significant amount of time, thus allowing for the general possibility of life --- although the habitability of moons and planets harbored by M-dwarfs is often seriously imperiled by highly energetic stellar activity \citep[e.g.,][]{france13,cuntz16,young17}.  Moreover, the need for a large exomoon to stabilize the climate may have been overstated as mild obliquity variations can occur over billion-year timescales without a moon \citep{Lissauer2012}.  Also, distant perturbers (i.e., companion stars) can endanger planets with moons through spin--orbit resonances \citep{Quarles2019,Quarles2022}.

The search for exomoons in M-dwarf systems has gained increasing prominence in recent years \citep{hek4, jwst24}. As a potential mode for habitability, exomoons around HZ planets may be an important piece of the exolife puzzle.  Hence, for future observations and research proposals, it would be pertinent to gauge the lifetimes of exomoons in relevant systems; moreover, exomoons themselves could potentially harbor life \citep[e.g.,][]{kalt10, heller14}.

Previous examples of theoretical work on exomoons include studies by \citet{quarles12}, \citet{Rosario-Franco2020}, \citet{jagtap21}, and \citet{patel25} that identify conditions under which exomoons could exist in distinct star--planet systems.  These studies identified general regions for moon stability as 3-body systems, including tidal effects and retrograde orbits, where perturbations from nearby planets are assumed to be negligible \citep{Payne2013}.

Other works exploring exomoon migration, including tidal effects, comprise \citet{barnes02}, \citet{sasaki12}, and \citet{piro18}. In \citet{barnes02}, the exomoons were found to migrate inward or outward, or to be tidally disrupted or stripped, depending on the system parameters, particularly the star--planet distance and the involved masses as previously shown in \citet{murraydermott1999}.
\citet{sasaki12} expanded on this work by including tides due to the Moon and Sun, and associated effects on planetary spin, which ultimately led to longer lifetimes for the exomoon.
\citet{piro18} then focused on Earth--Moon-like systems and found that the planet's initial spin was critical to the fate of the moon, with a small initial spin leading to the moon most likely being stripped. Additionally, the asynchronous spin of the planet due to tidal interactions led to Earth-like tidal heating for ${\sim}10^9$ yr.

The aim of this work is to examine the possibility of moons hosted by K2-18b using the N-body integrator \texttt{rebound} and the recently developed tidal evolution module \citep{tides_spin} for  \texttt{reboundx} \citep{reboundx}. Our work represents one of the first systematic efforts to assess exomoon stability in a habitable-zone system using the full capabilities of the reboundx tidal module, including spin evolution, allowing more realistic constraints than previous analytical estimates.  
Our paper is structured as follows.  In Section~\ref{sec:methods}, we discuss the methods adopted, notably aspects of the N-body integrator \texttt{rebound}.  Sections~\ref{sec:results} and \ref{sec:summary} report our results, and the summary and conclusions, respectively.

\section{Methods} \label{sec:methods}

We use the N-body integrator, \texttt{rebound} \texttt{(v4.4.3)} \citep{rebound}, along with the extension library, \texttt{reboundx} \texttt{(v4.3.0)} \citep{reboundx} to investigate the potential orbital stability of exomoons orbiting K2-18b with tidal effects. We include the \texttt{tides\_spin} module \citep{tides_spin}, which implements the constant time-lag model of tides \citep{eggleton}.

The simulations consist of a star, planet, and moon where the planet's mass $M_{\rm p}$, eccentricity $e_{\rm p}$, and time-lag $\tau_{\rm p}$ are varied within the 1$\sigma$ uncertainties reported in recent studies \citep{sarkis18,benneke19}. We vary planetary mass range from 7.28 -- 9.98 $M_{\oplus}$, stepping by 0.01 $M_{\oplus}$ (271 values). Three values for the initial planetary eccentricity $e_{\rm p}$ are used (0.12, 0.20, and 0.28), which represent the lower, median, and upper bound values established by observations \citep{sarkis18}; see Table~\ref{tab:init_con} for details. Additionally, we use three different time-lag $\tau_{\rm p}$ constants for the planet: $10$, $100$, and $698\ {\rm s}$, each representing a different potential planetary composition that influences the tidal dissipation, with $698~{\rm s}$ representing current Earth-like conditions \citep{neron97, bolmont15}. 

Overall we perform 2439 simulations, where we integrate over a $10^7$ yr timescale with an initial integration timestep equal to 20\% of the moon's initial period. The \texttt{TRACE} integrator \citep{Lu2024} is used along with the \texttt{IAS15} \citep{Rein2015} integrator for close approaches between the host planet and putative moon.

In each simulation, the putative exomoon is equal in mass to Earth's moon (i.e., Luna) and its orbit begins with a small non-zero eccentricity ($e_{\rm m}$) of $10^{-6}$ and an initial semi-major axis ($a_{\rm m}$) that is $3\times$ the Roche limit of a fluid satellite, calculated by

\begin{align}
    d \ = \ 2.44 R_{\rm p} \left(\frac{\rho_{\rm p}}{\rho_{\rm m}}\right)^{1/3}
\end{align}
and
\begin{align}
    \frac{\rho_{\rm p}}{\rho_\oplus} = \left(\frac{M_{\rm p}}{M_\oplus}\right)\left(\frac{R_{\rm p}}{R_\oplus}\right)^{-3}~,
\end{align}
where $M_{\rm p}$, $R_{\rm p}$, and $\rho_{\rm p}$ denote the mass, radius, and density of the planet, respectively, while $\rho_{\rm m}$ and $\rho_{\oplus}$ denote the density of the moon ($3.34\ {\rm g~cm^{-3}}$) and of the Earth ($5.515\ {\rm g~cm^{-3}}$).  In addition, $M_\oplus$ and $R_\oplus$ also indicate terrestrial values.

The planet is evolved assuming: (1) a constant radius of 2.61$R_\oplus$ \citep{benneke19}, in contrast to the host star and moon (which are treated as point masses), and (2) an initial spin period of 5 hours, which is representative of the initial spin period of protoplanets \citep{Kokubo2010,Takaoka2023}. We assume two Love numbers $k_{2}$ (0.298 and 0.120) and moment of inertia constants $\overline{C}$ (0.3308 and 0.2200), representing Earth-like and Neptune-like conditions with respect to $k_2$ and $\overline{C}$ only; see Table~\ref{tab:init_con} for details.

The exomoon in our simulation is defined as unstable once it passes the critical semi-major axis $a_{\rm crit}$ \citep{Rosario-Franco2020}
\begin{align}\label{tab:stablim}
    a_{\rm crit} \ = \ 0.4031 (1-1.123e_{\rm p}) R_{\rm H}~,
\end{align}
where $e_{\rm p}$ denotes the planetary eccentricity and $R_{\rm H}$ denotes the Hill radius of the planet. The maximum lifetime of the moon ($t_{\rm max}$ in yr), or the length of time until the migrating moon passes this stability limit, is tracked as a proxy for stability. 

\begin{table*}
	\centering
	\caption{Initial conditions for the \texttt{rebound} simulations of the K2-18 system. Data for the first 3 columns obtained from \citet{benneke19} and \citet{sarkis18} for $e_{\rm p}$. All symbols have their usual meaning except if denoted otherwise in the text.} 
	\label{tab:init_con}
	\begin{tabular}{lccccccccccccc}
		\hline
  		 & $M_{\ast}$ & $M_{\rm p}$ & $a_{\rm p}$ & $e_{\rm p}$ & $k_{\rm 2}$ & $\overline{C}$ & $P_{\rm s}$ & $\tau_{\rm p}$ & $M_{\rm m}$ & $a_{\rm m}$ & $e_{\rm m}$ \\
             & ($M_{\odot}$) & ($M_{\oplus}$) & (au) & & & & (h) & (s) & ($M_{\oplus}$) & (au) & \\
		\hline
        Earth-like & 0.4951 & 7.28--9.98 & 0.1591 & 0.12, 0.20, 0.28 & 0.298 & 0.3308 & 5 & 10, 100, 698 & 0.0123 & $3R_{\rm Roche}$ & $10^{-6}$\\
        Neptune-like & 0.4951 & 7.28--9.98 & 0.1591 & 0.12, 0.20, 0.28 & 0.120 & 0.2200 & 5 & 10, 100, 698 & 0.0123 & $3R_{\rm Roche}$ & $10^{-6}$\\
		\hline
	\end{tabular}
\end{table*}

\section{Results} \label{sec:results}

\begin{figure*}
	\includegraphics[scale=0.7]{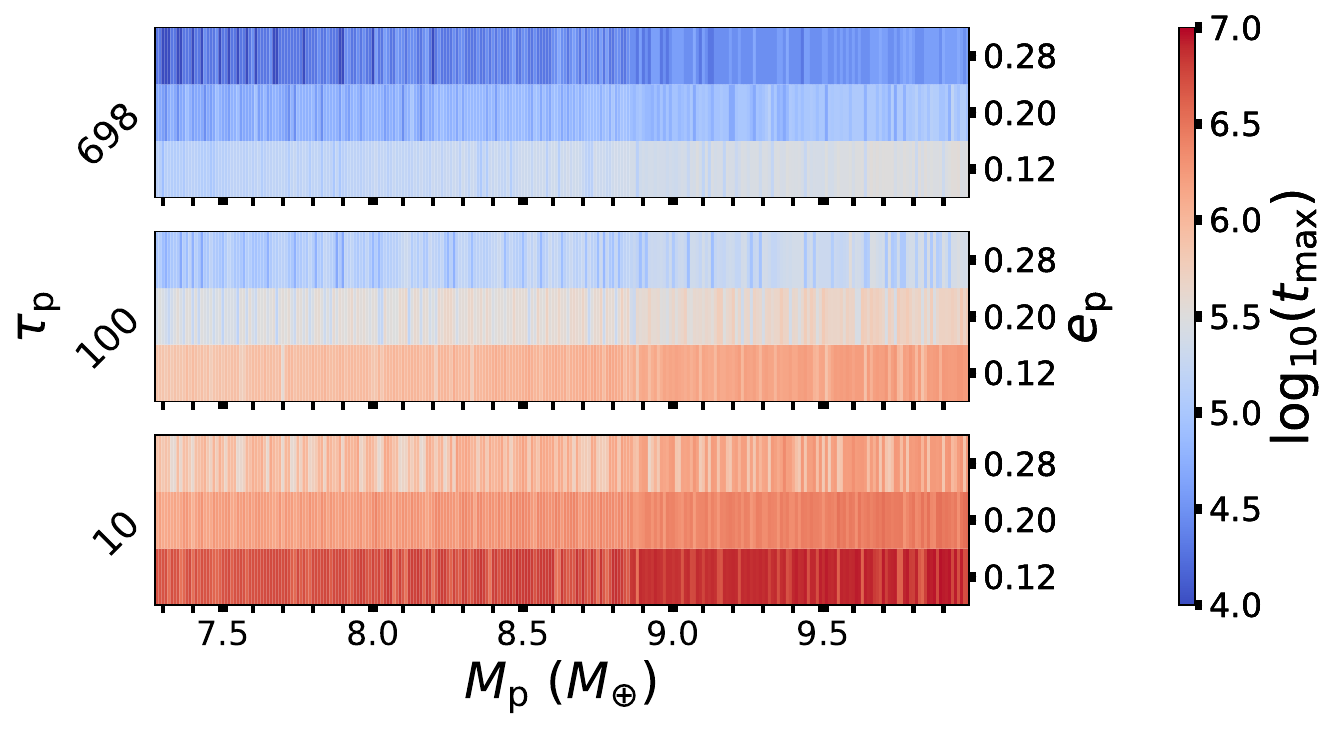}
    \caption{Logarithm of exomoon lifetime ($\rm log_{10}(\text{$t_{\rm max}$})$; color coded) from stability simulations that vary a host planet's mass, tidal constant time-lag, $\tau_{\rm p}$, and initial eccentricity while using Earth-like parameters for the $k_2$ Love number and moment of inertia constant, $\overline{C}$. }
    \label{fig:earth}
\end{figure*}

\begin{figure*}
	\includegraphics[scale=0.7]{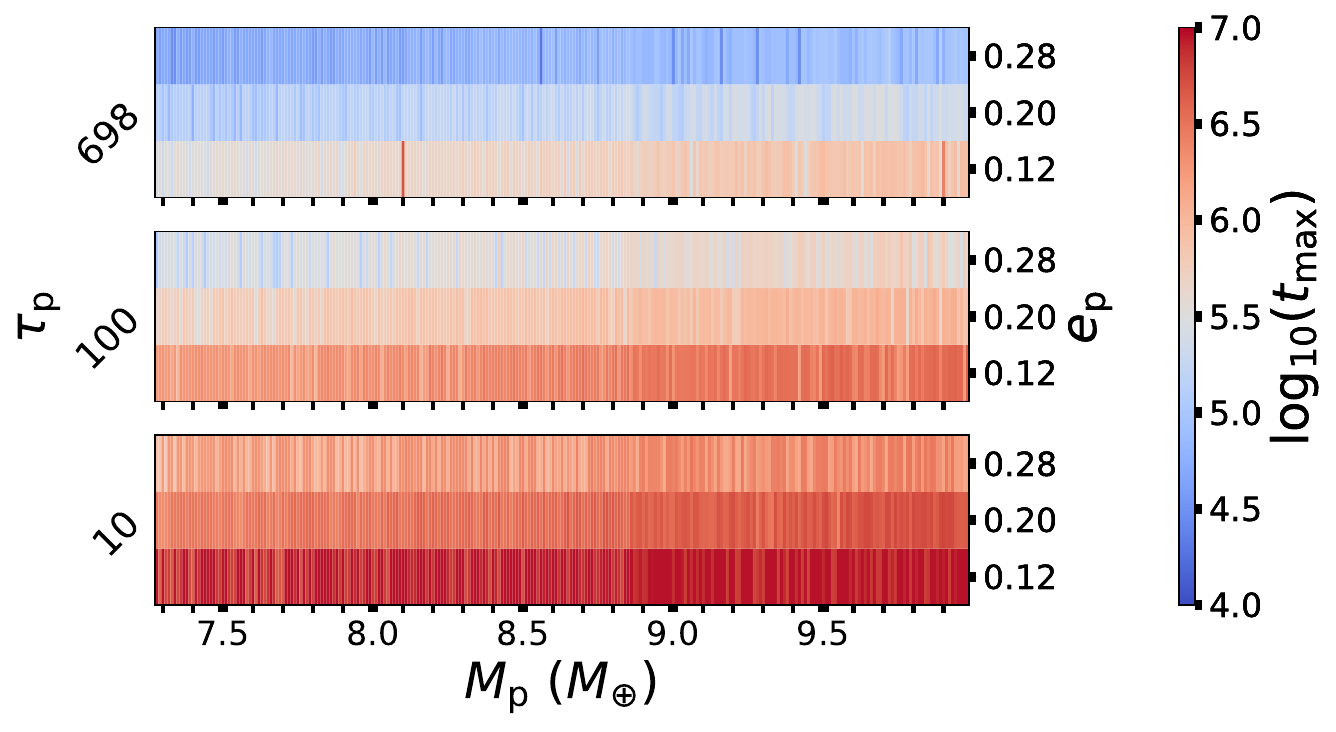}
    \caption{Same as Fig.~\ref{fig:earth}, but using Neptune-like parameters for $k_2$ and $\overline{C}$.}
    \label{fig:neptune}
\end{figure*}

The results of our numerical simulations are described by the lifetime of the moon as depicted through contour maps (see Figs.~\ref{fig:earth} and \ref{fig:neptune}) as a proxy for stability, with longer lifetimes ($10^7$ yr) represented by redder cells and shorter lifetimes ($10^4$ yr) by bluer cells. 

Figure~\ref{fig:earth} considers an Earth-like $k_2$ and $\overline{C}$, where the maximum lifetime of the exomoon from our numerical simulations ranges from $10^4$ to $10^7$ yr. This variance in maximum lifetime depends on the constant time lag parameter $\tau_{\rm p}$, where a value for $\tau_{\rm p}=698\ {\rm s}$ leads to the shortest maximum lifetimes (closer to the $10^4$ yr limit), whereas a value of $\tau_{\rm p} = 10\ {\rm s}$ to the longest lifetime, as expected given the quicker assumed dissipation and faster moon migration rate \citep{Hut1981,Barnes2017}.

The maximum lifetime of the putative exomoon depends on the initial planetary eccentricity $e_{\rm p}$, typically within a factor of ${\sim}3.16$, or $\sqrt{10}$. A higher $e_{\rm p}$ value leads to shorter moon lifetimes due to a stronger perturbation from the host star at the planet's pericenter, which increases the likelihood for ejection after the moon migrates past ${\sim}0.3\ R_{\rm H}$. \cite{Rosario-Franco2020} determined that higher $e_{\rm p}$ values lead to closer-in stability limits as seen in Eq.~\ref{tab:stablim}, thus requiring the moon to migrate a shorter distance to reach our instability criterion. We find that the maximum lifetime of the moon in all Earth-like cases (in terms of $k_2$ and $\overline{C}$) is 9.06 Myr.

Extending this approach to the Neptune-like case, we find very similar trends to the Earth-like case. These simulations reveal (in Fig. \ref{fig:neptune}) a similar range in maximum lifetime in the $10^4-10^7$ year range while also finding the same trends for higher $\tau_{\rm p}$ and $e_{\rm p}$ values. However, there is an overall shift toward longer lifetimes across the board in the Neptune-like case. A lower $k_2$ value leads to slower $a_{\rm m}$ and $e_{\rm m}$ evolution \citep{Hut1981,eggleton,Barnes2017}. 

For both cases, we find that the moon becomes orbitally unstable within ${\sim}9.1$ Myr even at lowest time-lag $\tau_{\rm p}= 10\ {\rm s}$ and planetary eccentricity $e_{\rm p}=0.12$ with Neptune-like parameters. Based on this timescale, no observable moons are expected to exist around K2-18b, considering that system's age is larger by about a factor of 300.

\begin{figure*}
	\includegraphics[scale=0.7]{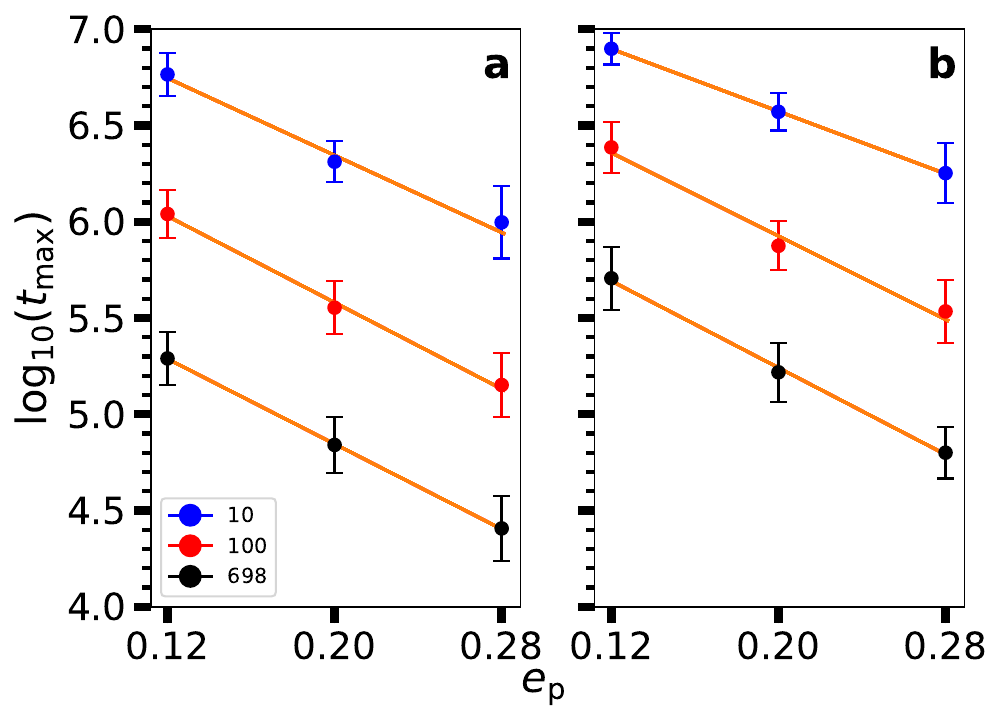}
    \caption{Median $\log_{10}(t_{\rm max})$ values with 1$\sigma$ error bars over the range of $M_{\rm p}$ values (rows in Figs. \ref{fig:earth} and \ref{fig:neptune}) for the (a) Earth-like and (b) Neptune-like results. Orange lines represent samples using Markov Chain Monte Carlo (MCMC) methods from \texttt{emcee}. Black, red, and blue points represent $\tau_{\rm p}$ values of 698, 100, and 10~s, respectively. }
    \label{fig:stats}
\end{figure*}

There are variations in the maximum lifetime for a given value in $\tau_{\rm p}$ and $e_{\rm p}$ (in Figs. \ref{fig:earth} and \ref{fig:neptune}) due to the chaotic nature of the 3-body problem.  Small changes in the assumed planetary mass proliferate different outcomes around the median value of the respective row. Figure~\ref{fig:stats} shows the median $\rm log_{10}(\text{$t_{\rm max}$})$ values and 1$\sigma$ standard deviations for each row in the (a) Earth-like and (b) Neptune-like plots. Time-lag $\tau_{\rm p}$ values of 698, 100, and 10 s are represented by black, red, and blue points, respectively. The orange lines represent Markov Chain Monte Carlo (MCMC) sampling using the package \texttt{emcee} \citep{goodman10, emcee13}. Given a set of three data points with error bars, \texttt{emcee} fits a model to the data and provides a best-fit line given these data points.

We find all three of the Earth-like lines and the $698$ and $100$ s Neptune-like lines to have similar slopes of $\sim -5$ to $-5.5$ while the $10$ s Neptune-like case has a distinct slope $m$ of $\sim -4$.  Additionally, we determine the $y$-intercepts $b$ of these best-fit lines, which represents the maximum lifetime (in $\log_{\rm 10}[{\rm yr}]$) of the moon in the limit of circular planetary eccentricity $(e_{\rm p}=0$; see Table \ref{tab:emcee}). In the circular planetary orbit limit (which corresponds to about 1.5 standard deviations in the $e_{\rm p}$ uncertainty if a Gaussian distribution for the errors is assumed), the increase in the maximum lifetime is less than an order of magnitude, where the longest lifetime is ${\sim}22-24$ Myr. This timescale is much lower than the system lifetime. The error bars represent the variation in the maximum lifetime, which resulted in variations $<0.5$ in $\log_{\rm10}$ and are not statistically significant. These results can act as a predictive filter for future exomoon surveys by identifying potential false positives and helping to prioritize future exomoon targets.

\begin{table}
	\centering
	\caption{Best-fit slope (m) and y-intercept (b) for the data in Fig.~\ref{fig:stats}, derived using \texttt{emcee} \citep{emcee13} for each $\tau$. The slope $m$ measures the dependence of the median lifetime $\log_{10}(t_{\rm max})$ on planet eccentricity. The y-intercept $b$ is the maximum lifetime for a circular orbit.}
	\label{tab:emcee}
	\begin{tabular}{lccc} 
    \hline
      		& $\tau_{\rm p}$ & $m$ & $b$ \\
            & (s) & ($\log_{10}[{\rm yr}]$) & ($\log_{10}[{\rm yr}]$)\\
		\hline
                 & 698 & $-$5.53 & 5.95 \\
      Earth-like & 100 & $-$5.61 & 6.70 \\
                 & 10 & $-$5.01 & 7.35 \\
        \hline
                   & 698 & $-$5.63 & 6.37 \\
      Neptune-like & 100 & $-$5.41 & 7.01 \\
                   & 10 & $-$4.05 & 7.38 \\
		\hline
	\end{tabular}
\end{table}


\section{Summary and Conclusions} \label{sec:summary}

The focus of this work is to examine the possibility of an exomoon hosted by K2-18b using the N-body integration package \texttt{rebound} and its extension library \texttt{reboundx} for additional tidal effects. We consider two main configurations for the planet, one with an Earth-like $k_2$ Love number and moment of inertia constant $\overline{C}$ and one with Neptune-like values.
We simulate 3-body star--planet--moon systems for $10^7$ yr while varying the planetary mass $M_{\rm p}$, the planetary eccentricity $e_{\rm p}$, and the planetary tidal time-lag $\tau_{\rm p}$. These simulations consider tides along with standard 3-body effects by using the \texttt{tides\_spin} implementation from \texttt{reboundx}, which is based on the constant time-lag model of tides \citep{eggleton}.

From these simulations, we find that while Neptune-like configurations in general have longer lifetimes than Earth-like ones, neither case has a lifetime longer than $10^7$ yr. Small changes in the K2-18b's spin are sufficient to migrate a putative moon beyond the stability limit \citep{Rosario-Franco2020} due to the relatively small Hill radius.  This migration timescale is short enough that a migration reversal \citep[e.g.,][]{sasaki12} is not possible, assuming that the K2-18b begins as a rapid rotator.  The moon's lifetime is low compared to the stellar age (and that of the planet), given as close to 3~Gyr.  Hence, it can be safely concluded that this system lacks observable moons. Although the simulations consider only a single moon mass, we note that varying the moon’s mass would have a limited effect on its survival time, modifying the lifetime by no more than a factor of a few. Given K2-18b’s controversial classification as a potentially Hycean world and the recent JWST observations suggesting atmospheric biosignatures, the lack of stable moons would eliminate a key avenue for additional habitable environments in this system—moons that could stabilize climate, host subsurface oceans, or even present separate biosignature targets. 


\section*{Acknowledgements}

    S.D.P. and N.N.W. acknowledge support by the National Science Foundation (NSF) under grant No. AST-2054353.  B.Q. acknowledges support in part by the Texas A\&M High Performance Research Computing (HPRC) and the NSF under grant No. 2232895. The authors acknowledge the Texas A\&M HPRC for providing computing resources on the Launch cluster that contributed to the research results reported here.

\section*{Data Availability}

The data underlying this article will be shared on reasonable request to the corresponding author.



\bibliographystyle{mnras}
\bibliography{k218} 








\bsp	
\label{lastpage}
\end{document}